\documentclass[aps,pra,superscriptaddress,twocolumn,showkeys]{revtex4-2}

\usepackage[T1]{fontenc}         
\usepackage[utf8]{inputenc}        
\usepackage{lmodern}               
\usepackage{graphicx} 
\usepackage{amssymb}
\usepackage{amsmath}
\usepackage{underscore}
\usepackage{xcolor}

\usepackage{xcolor} 

\begin{document}

\title{High-Dimensional Deterministic Secure Quantum Communication with Reed-Solomon Erasure Coding}

\author{L. F. A. de Sousa Moura}
\email{luiz.f.s.m@posgrad.ufsc.br}
\affiliation{Department of Informatics and Statistics, Federal University of Santa Catarina, CEP 88040-900, Florian\'{o}plis, SC, Brazil}

\author{G. L. Zanin}
\email{guilherme_zanin@ufg.br}
\affiliation{Department of Physics, Federal University of Goias, CEP 74690-900, Goiania, GO, Brazil}

\author{P. H. Souto Ribeiro}
\email{p.h.s.ribeiro@ufsc.br}
\affiliation{Department of Physics, Federal University of Santa Catarina, CEP 88040-900, Florian\'{o}plis, SC, Brazil}

\author{C. Becker Westphall}
\email{carlos.westphall@ufsc.br}
\affiliation{Department of Informatics and Statistics, Federal University of Santa Catarina, CEP 88040-900, Florian\'{o}plis, SC, Brazil}

\begin{abstract}
Deterministic Secure Quantum Communication (DSQC) is a quantum cryptographic technique engineered to transfer a message through a quantum channel, requiring an auxiliary classical channel for eavesdropping verification and decoding, but without prior key distribution. This article presents a theoretical high-dimensional prepare and measure DSQC protocol using the Reed-Solomon erasure coding to ensure data resilience to noise. This protocol offers the following benefits: it eliminates the need for quantum memory or entanglement, it can be built with commercially available technology, and its higher capacity improves the overall transmission rate.
\end{abstract}

\keywords{Deterministic Secure Quantum Communication, Quantum Cryptography, Erasure Coding, Quantum Optics, Fourier Optics}

\maketitle

\section{Introduction}
Much of the current Internet will become vulnerable with the advent of a sufficiently powerful quantum computer. Not just the message and the key exchange can be intercepted by Man-in-the-Middle (MitM) attacks, but also important packets can be stored now to be decrypted years ahead, in an attack named Store-Now-Decrypt-Later (SNDL)~\cite{b1,b2,b3}. An alternative to mitigate this threat is using quantum cryptography, which is a field that explores the principles of quantum mechanics in favor of cryptography.

Since the seminal publication of 1984~\cite{b4}, the first Quantum Key Distribution (QKD) protocol, and later the Shor's algorithm~\cite{b5}, an algorithm that theoretically breaks most of the current classical key distribution schemes, quantum cryptography has focused efforts on this class of protocols. However, the properties of quantum physics allow not just the exchange of keys but the message itself can be encoded into quantum states and safely transferred through the network, in a new class of protocols called Quantum Secure Direct Communication (QSDC). The first QSDC protocol was proposed in 2002~\cite{b6}.

QSDC is defined as the technique to safely transfer a message through quantum states. There is an alternative class of protocols that is also engineered for the same problem, but requires an auxiliary transfer of classical bits for a successful decoding on the receiver's side. This class of protocols is named Deterministic Secure Quantum Communication (DSQC).

Both QSDC and DSQC suffer from low bit rate. This is a more immediate concern for these classes of protocols than it is for QKD, since QKD is restricted to transmitting the key and not the entire message. But a common form of mitigating this concern is to use higher dimensionality. Quantum communication is also a victim of network noise, which causes loss of information in transit. As a mitigation, some authors propose a fault-tolerant approach for QSDC and DSQC~\cite{b7,b8,b9,b10,b11}.

In this article, a novel DSQC protocol is proposed, combining classical erasure coding with quantum prepare and measure qudit exchange. By leveraging Reed-Solomon (RS) code~\cite{b12}, it is possible to transfer a deterministic message through a noisy network. The proposed scheme has the advantages of avoiding the use of quantum memory and can be implemented with available technology. We also present a proposal for the practical implementation using single photons and the Transverse Spatial Degrees of Freedom (TSDF) of light. This work potentially opens new avenues towards DSQC with quantum structured light. 

\section{Reed-Solomon based DSQC secured by BB84 non cloning strategy}

\begin{figure}[!htb]
    \centering
    \includegraphics[width=1\linewidth]{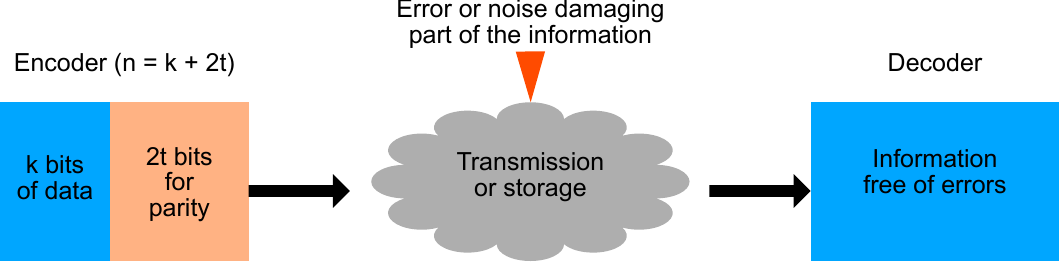}
    \caption{Reed-Solomon}
    \label{fig:reed-solomon}
\end{figure}

We propose the implementation of a DSQC communication scheme using the platform introduced in Ref.~\cite{b101}, where a QKD scheme was demonstrated. It used the TSDF of light to encode more than one bit of information per transmitted photon. We adapt it in order to share not just a random string of characters between two parties, but also a deterministic message with the help of the RS erasure coding. See Fig.~\ref{fig:reed-solomon} for a block diagram. The sender appends parity bits to the message bits, making it resistant to errors. This means that even if part of the information is lost in transit, the message can still be recovered at the receptor. Along with the RS, we include preparation and measurements in Mutually Unbiased Basis (MUB) in the same way as in the BB84 protocol, used to protect against intercept resend attacks. 

Let us illustrate the scheme with a simple example.  Suppose that Alice has a secret message consisting of $k = 4$  characters and she wants to send the message secretly to Bob, protected from the eavesdropper Eve. Alice prepares the message by choosing randomly between two MUBs and Bob also performs measurements in one of two MUBs randomly selected. Anticipating that in half of the measurements Bob will choose the wrong basis and will not be able to recover the information properly, Alice sends extra characters. They will be used to test the security of the channel (decoy states) and to implement the RS strategy (parity characters). She will use an optical communication channel with single photon carriers encoded with $qudits$ of dimension $d = 11$. In the present example she sends a total of 10 single photons (4 representing the message and 6 parity photons). Each contains the information appropriated to the size of the alphabet used.

\begin{table}[!htb]
    \centering
    \begin{tabular}{|ccccccccccc|}
    \hline
         O& A& E& D& S& T& J& V& H& R& _\\
         0& 1& 2& 3& 4& 5& 6& 7& 8& 9& 10\\
    \hline
    \end{tabular}
    \caption{Simple alphabet}
    \label{tab:alphabet}
\end{table}

According to RS, the alphabet must have a size equal to a prime number $q$. In this example, the alphabet size is $q=11$ as shown in Table~\ref{tab:alphabet}. Suppose that Alice wants to send the word DATA to Bob. According to the chosen alphabet, this word corresponds to the numbers 3-1-5-1 in this order. 

The numerical sequence representing the message to be transferred is then transformed by Alice into a polynomial of degree $k-1$:

\begin{equation}
\label{eq:polynomial}
    P(x)=3+1 \cdot x^1 +5 \cdot x^2+1 \cdot x^3.
\end{equation}

Alice proceeds by evaluating the polynomial at $n=10$ positions, where $n$ is the number of photons to be sent:

\begin{equation}\label{eq:polynomialseval}
\begin{aligned}
P(1) &= 3 + 1 \cdot 1^1 + 5 \cdot 1^2 + 1 \cdot 1^3 \equiv 10 \pmod{11} \\
P(2) &= 3 + 1 \cdot 2^1 + 5 \cdot 2^2 + 1 \cdot 2^3 \equiv 0 \pmod{11} \\
P(3) &= 3 + 1 \cdot 3^1 + 5 \cdot 3^2 + 1 \cdot 3^3 \equiv 1 \pmod{11} \\
P(4) &= 3 + 1 \cdot 4^1 + 5 \cdot 4^2 + 1 \cdot 4^3 \equiv 8 \pmod{11} \\
P(5) &= 3 + 1 \cdot 5^1 + 5 \cdot 5^2 + 1 \cdot 5^3 \equiv 5 \pmod{11} \\
P(6) &= 3 + 1 \cdot 6^1 + 5 \cdot 6^2 + 1 \cdot 6^3 \equiv 9 \pmod{11} \\
P(7) &= 3 + 1 \cdot 7^1 + 5 \cdot 7^2 + 1 \cdot 7^3 \equiv 4 \pmod{11} \\
P(8) &= 3 + 1 \cdot 8^1 + 5 \cdot 8^2 + 1 \cdot 8^3 \equiv 7 \pmod{11} \\
P(9) &= 3 + 1 \cdot 9^1 + 5 \cdot 9^2 + 1 \cdot 9^3 \equiv 2 \pmod{11} \\
P(10) &= 3 + 1 \cdot 10^1 + 5 \cdot 10^2 + 1 \cdot 10^3 \equiv 6 \pmod{11}
\end{aligned}
\end{equation}

The results are values modulo $q$ and they form the codeword $C=\{10, 0, 1, 8, 5, 9, 4, 7, 2, 6\}$ that Alice will send to Bob. She must also include decoy photons at random positions in order to calculate the Quantum Bit Error Rate (QBER) and evaluate if Eve is performing intercept-resend attacks: $C_d=\{\textbf{1}, \textbf{7}, 10, \textbf{4}, 0, \textbf{8}, 1, \textbf{5}, \textbf{3}, 8, 5, \textbf{10}, 9, \textbf{2}, 4,$ $\textbf{9}, 7, \textbf{8}, 2, 6, \textbf{1}\}$, where the numbers in boldface are decoys, positioned at random positions and assuming random values between $0$ and $q-1$. 

Alice encodes the characters of $C_d$ in the TSDF of light. The preparation is made in one of two MUBs and the photons are sent sequentially with a timestamp. After sending the photons through the quantum channel, she uses the authenticated classical channel to announce the positions and values of the decoy states.

Bob receive the photons and measures them in one of two MUBs. He also receives the information about the decoys and evaluates the QBER in order to decide if the connection is free from intercept-resend eavesdropping. If the connection is not secure, the communication is interrupted. If the connection is secure, Alice informs the measurement bases for the non-decoy photons. The ordering of the photons is also updated by eliminating the positions of the decoy photons. 

Bob receives a fraction of $C$, considering that part of the photons were measured in the wrong basis and part of the photons were lost. If Bob has at least $k$ valid photons, he can successfully read the original message. Let us suppose that he has $k=4$ photons left at the end of the procedure. Now he has a system of four equations and has to find four variables:

\begin{equation}
\label{eq:polynomialsbob}
\begin{aligned}
P(3) &= c_0 + c_1 \cdot 3^1 + c_2 \cdot 3^2 + c_3 \cdot 3^3 \equiv 1 \pmod{11} \\
P(5) &= c_0 + c_1 \cdot 5^1 + c_2 \cdot 5^2 + c_3 \cdot 5^3 \equiv 5 \pmod{11} \\
P(6) &= c_0 + c_1 \cdot 6^1 + c_2 \cdot 6^2 + c_3 \cdot 6^3 \equiv 9 \pmod{11} \\
P(9) &= c_0 + c_1 \cdot 9^1 + c_2 \cdot 9^2 + c_3 \cdot 9^3 \equiv 2 \pmod{11}
\end{aligned}
\end{equation}

All he has to do is solve the system of equations to retrieve the message as $c_0 = 3$, $c_1 = 1$, $c_2 = 5$, $c_3 = 1$. It is important to notice that the RS scheme allows that redundant information is transmitted without weakening the security. 

\subsection{Splitting the transmission into blocks}

RS imposes a limitation on the length of the message to be transferred. In a $q$ dimensional alphabet, all polynomials are evaluated modulo $q$ and it is not possible to transfer a number $n > q$ of photons per transmission. Since $x = 0$ results in the trivial result $P(0) = c_0$, this option is usually discarded, and the total number of photons to be transmitted in a single block should be $n < q$.

Let us now suppose that Alice wants to transfer a list of words to Bob and this list of words is longer than the size of the alphabet $q = 11$. As an example, the string JADE\_STORED\_HER\_DATA is 20 characters long. Using the same number of four characters per block and the remaining characters reserved for parity, the string can be split into five blocks. The five blocks that represent the original string are \{JADE, \_STO, RED\_, HER\_, DATA\} and are numerically represented as \{6-1-3-2, 10-4-5-0, 9-2-3-10, 8-2-9-10, 3-1-5-1\}. Now, the five polynomials can be generated (Equation~\ref{eq:polynomials_blocks}).

\begin{equation}\label{eq:polynomials_blocks}
\begin{aligned}
P_1(x)=6+1 \cdot x^1 +3 \cdot x^2+2 \cdot x^3\\
P_2(x)=10+4 \cdot x^1 +5 \cdot x^2+0 \cdot x^3\\
P_3(x)=9+2 \cdot x^1 +3 \cdot x^2+10 \cdot x^3\\
P_4(x)=8+2 \cdot x^1 +9 \cdot x^2+10 \cdot x^3\\
P_5(x)=3+1 \cdot x^1 +5 \cdot x^2+1 \cdot x^3
\end{aligned}
\end{equation}

Alice proceeds by evaluating the polynomials $P_1(x)$, $P_2(x)$, $P_3(x)$, $P_4(x)$, $P_5(x)$ at the ten values of $x$ (Equations~\ref{eq:polynomialseval1_blocks},~\ref{eq:polynomialseval2_blocks},~\ref{eq:polynomialseval3_blocks},~\ref{eq:polynomialseval4_blocks},~\ref{eq:polynomialseval5_blocks}), corresponding to the ten photons sent per block.

\begin{equation}\label{eq:polynomialseval1_blocks}
\begin{aligned}
P_1(1) = 6 + 1 \cdot 1^1 + 3 \cdot 1^2 + 2 \cdot 1^3 \equiv 1 \pmod{11} \\
P_1(2) = 6 + 1 \cdot 2^1 + 3 \cdot 2^2 + 2 \cdot 2^3 \equiv 3 \pmod{11} \\
P_1(3) = 6 + 1 \cdot 3^1 + 3 \cdot 3^2 + 2 \cdot 3^3 \equiv 2 \pmod{11} \\
P_1(4) = 6 + 1 \cdot 4^1 + 3 \cdot 4^2 + 2 \cdot 4^3 \equiv 10 \pmod{11} \\
P_1(5) = 6 + 1 \cdot 5^1 + 3 \cdot 5^2 + 2 \cdot 5^3 \equiv 6 \pmod{11} \\
P_1(6) = 6 + 1 \cdot 6^1 + 3 \cdot 6^2 + 2 \cdot 6^3 \equiv 2 \pmod{11} \\
P_1(7) = 6 + 1 \cdot 7^1 + 3 \cdot 7^2 + 2 \cdot 7^3 \equiv 10 \pmod{11} \\
P_1(8) = 6 + 1 \cdot 8^1 + 3 \cdot 8^2 + 2 \cdot 8^3 \equiv 9 \pmod{11} \\
P_1(9) = 6 + 1 \cdot 9^1 + 3 \cdot 9^2 + 2 \cdot 9^3 \equiv 0 \pmod{11} \\
P_1(10) = 6 + 1 \cdot 10^1 + 3 \cdot 10^2 + 2 \cdot 10^3 \equiv 6 \pmod{11}
\end{aligned}
\end{equation}

\begin{equation}\label{eq:polynomialseval2_blocks}
\begin{aligned}
P_2(1) = 10 + 4 \cdot 1^1 + 5 \cdot 1^2 + 0 \cdot 1^3 \equiv 8 \pmod{11} \\
P_2(2) = 10 + 4 \cdot 2^1 + 5 \cdot 2^2 + 0 \cdot 2^3 \equiv 5 \pmod{11} \\
P_2(3) = 10 + 4 \cdot 3^1 + 5 \cdot 3^2 + 0 \cdot 3^3 \equiv 1 \pmod{11} \\
P_2(4) = 10 + 4 \cdot 4^1 + 5 \cdot 4^2 + 0 \cdot 4^3 \equiv 7 \pmod{11} \\
P_2(5) = 10 + 4 \cdot 5^1 + 5 \cdot 5^2 + 0 \cdot 5^3 \equiv 1 \pmod{11} \\
P_2(6) = 10 + 4 \cdot 6^1 + 5 \cdot 6^2 + 0 \cdot 6^3 \equiv 5 \pmod{11} \\
P_2(7) = 10 + 4 \cdot 7^1 + 5 \cdot 7^2 + 0 \cdot 7^3 \equiv 8 \pmod{11} \\
P_2(8) = 10 + 4 \cdot 8^1 + 5 \cdot 8^2 + 0 \cdot 8^3 \equiv 10 \pmod{11} \\
P_2(9) = 10 + 4 \cdot 9^1 + 5 \cdot 9^2 + 0 \cdot 9^3 \equiv 0 \pmod{11} \\
P_2(10) = 10 + 4 \cdot 10^1 + 5 \cdot 10^2 + 0 \cdot 10^3 \equiv 0 \pmod{11}
\end{aligned}
\end{equation}

\begin{equation}\label{eq:polynomialseval3_blocks}
\begin{aligned}
P_3(1) = 9 + 2 \cdot 1^1 + 3 \cdot 1^2 + 10 \cdot 1^3 \equiv 2 \pmod{11} \\
P_3(2) = 9 + 2 \cdot 2^1 + 3 \cdot 2^2 + 10 \cdot 2^3 \equiv 6 \pmod{11} \\
P_3(3) = 9 + 2 \cdot 3^1 + 3 \cdot 3^2 + 10 \cdot 3^3 \equiv 4 \pmod{11} \\
P_3(4) = 9 + 2 \cdot 4^1 + 3 \cdot 4^2 + 10 \cdot 4^3 \equiv 1 \pmod{11} \\
P_3(5) = 9 + 2 \cdot 5^1 + 3 \cdot 5^2 + 10 \cdot 5^3 \equiv 2 \pmod{11} \\
P_3(6) = 9 + 2 \cdot 6^1 + 3 \cdot 6^2 + 10 \cdot 6^3 \equiv 1 \pmod{11} \\
P_3(7) = 9 + 2 \cdot 7^1 + 3 \cdot 7^2 + 10 \cdot 7^3 \equiv 3 \pmod{11} \\
P_3(8) = 9 + 2 \cdot 8^1 + 3 \cdot 8^2 + 10 \cdot 8^3 \equiv 2 \pmod{11} \\
P_3(9) = 9 + 2 \cdot 9^1 + 3 \cdot 9^2 + 10 \cdot 9^3 \equiv 3 \pmod{11} \\
P_3(10) = 9 + 2 \cdot 10^1 + 3 \cdot 10^2 + 10 \cdot 10^3 \equiv 0 \pmod{11}
\end{aligned}
\end{equation}

\begin{equation}\label{eq:polynomialseval4_blocks}
\begin{aligned}
P_4(1) = 8 + 2 \cdot 1^1 + 9 \cdot 1^2 + 10 \cdot 1^3 \equiv 7 \pmod{11} \\
P_4(2) = 8 + 2 \cdot 2^1 + 9 \cdot 2^2 + 10 \cdot 2^3 \equiv 7 \pmod{11} \\
P_4(3) = 8 + 2 \cdot 3^1 + 9 \cdot 3^2 + 10 \cdot 3^3 \equiv 2 \pmod{11} \\
P_4(4) = 8 + 2 \cdot 4^1 + 9 \cdot 4^2 + 10 \cdot 4^3 \equiv 8 \pmod{11} \\
P_4(5) = 8 + 2 \cdot 5^1 + 9 \cdot 5^2 + 10 \cdot 5^3 \equiv 8 \pmod{11} \\
P_4(6) = 8 + 2 \cdot 6^1 + 9 \cdot 6^2 + 10 \cdot 6^3 \equiv 7 \pmod{11} \\
P_4(7) = 8 + 2 \cdot 7^1 + 9 \cdot 7^2 + 10 \cdot 7^3 \equiv 10 \pmod{11} \\
P_4(8) = 8 + 2 \cdot 8^1 + 9 \cdot 8^2 + 10 \cdot 8^3 \equiv 0 \pmod{11} \\
P_4(9) = 8 + 2 \cdot 9^1 + 9 \cdot 9^2 + 10 \cdot 9^3 \equiv 4 \pmod{11} \\
P_4(10) = 8 + 2 \cdot 10^1 + 9 \cdot 10^2 + 10 \cdot 10^3 \equiv 5 \pmod{11}
\end{aligned}
\end{equation}

\begin{equation}\label{eq:polynomialseval5_blocks}
\begin{aligned}
P_5(1) = 3 + 1 \cdot 1^1 + 5 \cdot 1^2 + 1 \cdot 1^3 \equiv 10 \pmod{11} \\
P_5(2) = 3 + 1 \cdot 2^1 + 5 \cdot 2^2 + 1 \cdot 2^3 \equiv 0 \pmod{11} \\
P_5(3) = 3 + 1 \cdot 3^1 + 5 \cdot 3^2 + 1 \cdot 3^3 \equiv 1 \pmod{11} \\
P_5(4) = 3 + 1 \cdot 4^1 + 5 \cdot 4^2 + 1 \cdot 4^3 \equiv 8 \pmod{11} \\
P_5(5) = 3 + 1 \cdot 5^1 + 5 \cdot 5^2 + 1 \cdot 5^3 \equiv 5 \pmod{11} \\
P_5(6) = 3 + 1 \cdot 6^1 + 5 \cdot 6^2 + 1 \cdot 6^3 \equiv 9 \pmod{11} \\
P_5(7) = 3 + 1 \cdot 7^1 + 5 \cdot 7^2 + 1 \cdot 7^3 \equiv 4 \pmod{11} \\
P_5(8) = 3 + 1 \cdot 8^1 + 5 \cdot 8^2 + 1 \cdot 8^3 \equiv 7 \pmod{11} \\
P_5(9) = 3 + 1 \cdot 9^1 + 5 \cdot 9^2 + 1 \cdot 9^3 \equiv 2 \pmod{11} \\
P_5(10) = 3 + 1 \cdot 10^1 + 5 \cdot 10^2 + 1 \cdot 10^3 \equiv 6 \pmod{11}
\end{aligned}
\end{equation}

The five polynomials evaluated at the ten points each will generate five codewords: $C_1 = \{1, 3, 2, 10, 6, 2, 10, 9, 0, 6\}$, $C_2 = \{8, 5, 1, 7, 1, 5, 8, 10, 0, 0\}$, $C_3 = \{2, 6, 4, 1, 2, 1, 3, 2, 3, 0\}$, $C_4 = \{7, 7, 2, 8, 8, 7, 10, 0, 4, 5\}$, $C_5 = \{10, 0, 1,$ $8, 5, 9, 4, 7, 2, 6\}$. Alice can now send the message through five communications or concatenate the five codewords and send them in a single transmission.

\subsection{Preparing a precise number of photons} 

In the example above, each message or block of message prepared $n = 10$ photons for each $k = 4$ characters of the portion of the message to be transmitted. A more precise way of measuring the number $n$ is to consider that the number of photons that Bob successfully measures follows a binomial distribution with mean value $\mu = n/2$. Simply sending $n = 2k$ photons still results in an unsuccessful reconstruction of the message half the times. Alice has to prepare a number of photons that satisfy the mean value being the number $k$ plus a statistical advantage (Equation~\ref{eq:statisticaladv}).

\begin{equation}\label{eq:statisticaladv}
    \mu = k + Z\sigma \implies \frac{n}{2} = k + Z \frac{\sqrt{n}}{2}
\end{equation}

Manipulating Equation~\ref{eq:statisticaladv}, for a given message length $k$, Alice has to prepare a number of photons according to Equation~\ref{eq:n}.

\begin{equation}\label{eq:n}
    n = \left\lceil\left(\frac{Z+\sqrt{Z^2+8k}}{2}\right)^2\right\rceil
\end{equation}

If the message is divided into blocks, the total number of photons per block should not exceed the length of the alphabet minus one. Since $n = q - 1$ is the maximum amount of photons she can transfer per block, she has to calculate the optimal number of characters $k$ in each block (Equation~\ref{eq:k}).

\begin{equation}\label{eq:k}
    k = \left\lfloor \frac{n}{2} - Z \frac{\sqrt{n}}{2}\right\rfloor
\end{equation}

In both Equations~\ref{eq:n} and~\ref{eq:k}, Alice can control the variable number $Z$ to optimize the chance of a successful communication versus the total number of photons to be transferred.

\section{Experimental implementation}

\begin{figure}[!htb]
    \centering
    \includegraphics[width=1\linewidth]{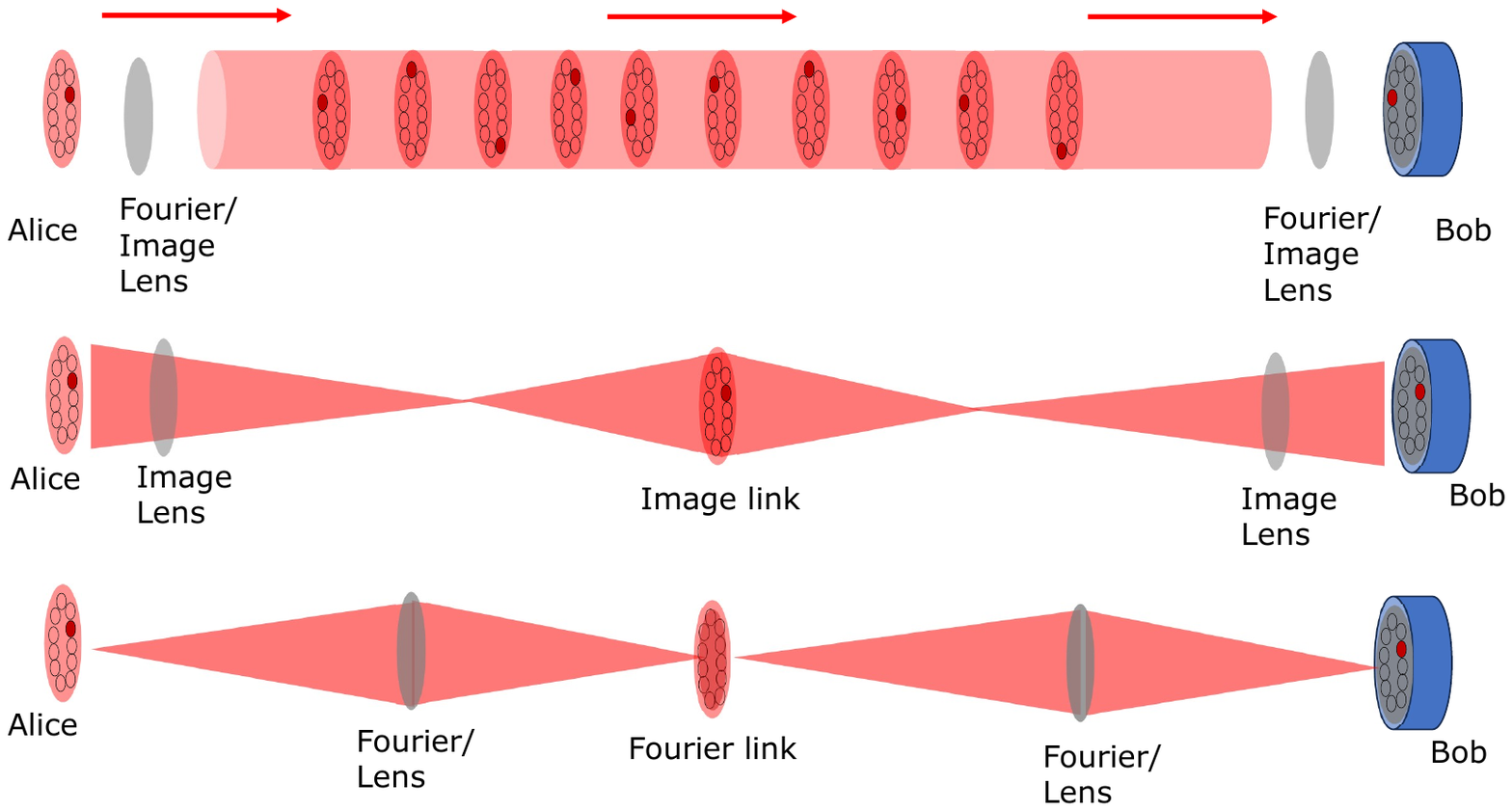}
    \caption{DSQC with Reed-Solomon protocol. Successful transmission with preparation and measurement in equal bases.}
    \label{scheme1}
\end{figure}
\begin{figure}[!htb]
    \centering
    \includegraphics[width=1\linewidth]{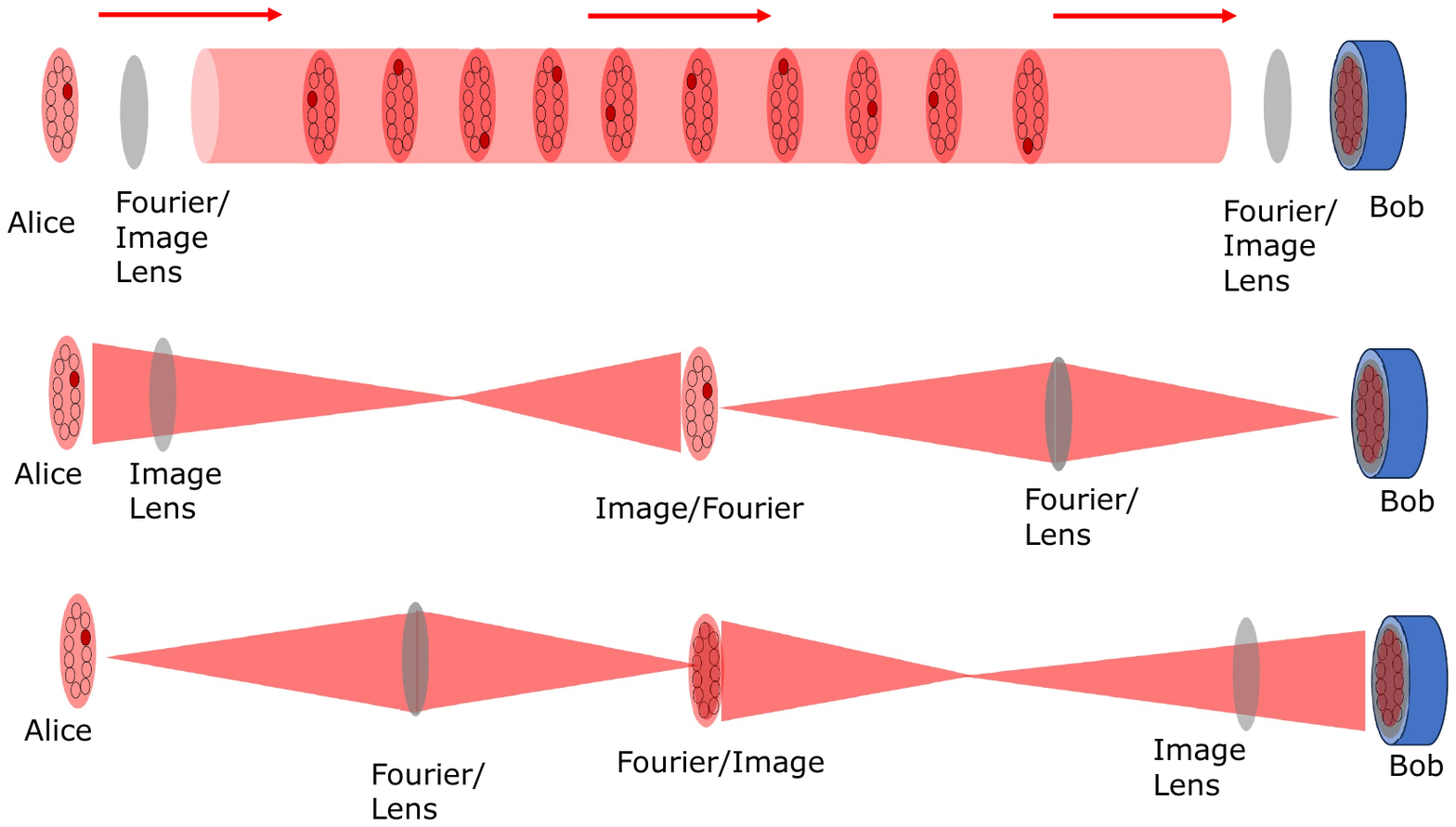}
    \caption{DSQC with Reed-Solomon protocol. Unsuccessful transmission with preparation and measurement in different bases.}
    \label{scheme2}
\end{figure}

In this section, we demonstrate that the QKD scheme introduced by Walborn et al.~\cite{b101} can be adapted to implement the RS-DSQC communication protocol described above. The protocol encodes both the message and the control parameters into the TSDF of a light beam. Using a Spatial Light Modulator (SLM), the transverse amplitude of the optical field is divided into spatial cells, where the position of each cell represents one symbol of a $d$-dimensional alphabet.

As illustrated in Figs.~\ref{scheme1} and \ref{scheme2}, the information is encoded into single-photon states distributed over a $d$-dimensional spatial alphabet. Alice and Bob publicly agree on the alphabet size $d$. In the example shown in the figure, we use $d = 11$. 

To transmit a symbol, Alice uses an SLM to generate a hologram in which the desired symbol is encoded into a specific diffracted cell. Light incident on the remaining regions of the SLM is simply reflected and discarded. After selecting the symbol to be transmitted, Alice randomly chooses between two MUBs: the position basis (near field) and the momentum basis (far field). The position basis is implemented by imaging the SLM through a $4f$ optical system, where a lens with focal length $f$ is placed at a distance $2f$ from both the SLM and the preparation plane, which serves as the reference plane for Bob. Similarly, the momentum basis is obtained by Fourier transforming the transverse field using a $2f$ configuration, with the lens positioned at a distance $f$ from both the SLM and the reference plane.

At the receiver, Bob independently chooses the basis used for the measurement. When Alice and Bob select the same basis, Bob recovers the symbol originally prepared by Alice. In the image basis, Alice images the SLM onto the preparation plane, while Bob images this plane onto his detector, reproducing the original transverse position as illustrated in the intermediate panel of Fig.~\ref{scheme1}. In the momentum basis, Alice performs the Fourier transform of the SLM plane, and Bob applies the corresponding inverse Fourier transform, recovering the symbol encoded by Alice, as illustrated in the bottom panel of Figure~\ref{scheme1}.

When Alice and Bob choose different bases, however, the detected intensity is spread over the array detector, forming a broad Gaussian-like distribution. In this case, Bob gains no information about the transmitted symbol, as shown in Fig.~\ref{scheme2}. Notice that the intensity profile is blurred at Bob's detection plane, meaning that the probability of detecting the photon is equal in all detection cells.  The same effect occurs if an eavesdropper intercepts the photon and performs the measurement in the wrong basis. Up to this point, the protocol follows the same operating principle as the QKD implementation of Ref.~\cite{b101}. The key difference is that the transmitted symbols are no longer random but are generated according to the RS protocol and decoded accordingly.

Compared with the implementation of Ref.~\cite{b101}, the proposed protocol introduces two main modifications. First, instead of single-pixel Avalanche Photodiodes (APDs), the detection stage can be implemented using single-photon CCD cameras or matrix APDs, as the ones in~\cite{b103}. In the case of matrix APDs, the optical system can be aligned such that each region of the hologram is mapped onto a specific detector pixel. Second, the protocol employs heralded single-photon states generated through Spontaneous Parametric down-Conversion (SPDC). One photon of the pair serves as the herald, while its twin propagates through the communication channel. The heralding event provides Bob with a precise timing reference and ensures that the detection events are correlated, thereby improving the security of the protocol.

\subsection{Experimental setup with heralded photons}

\begin{figure}[!htb]
    \centering
    \includegraphics[width=1\linewidth]{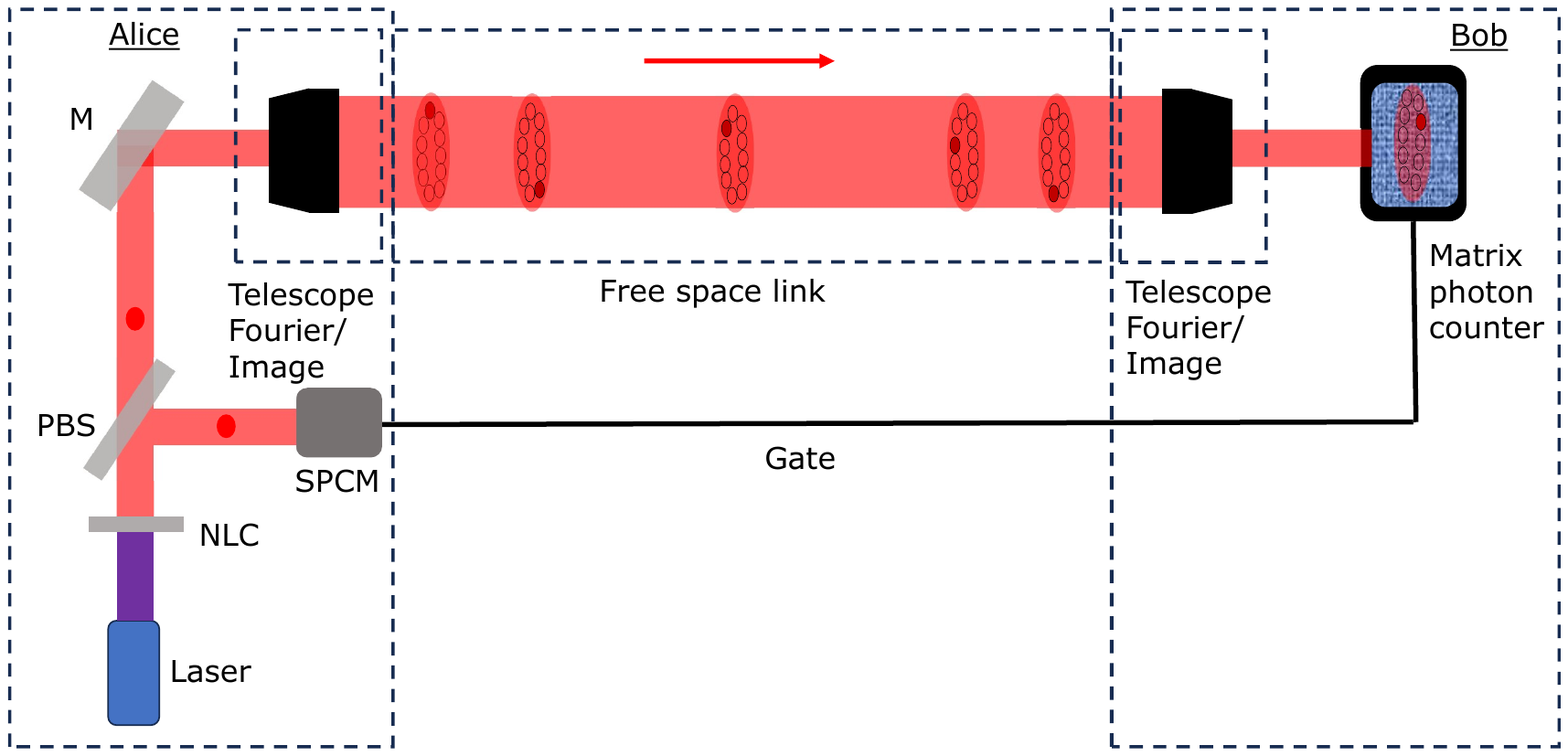}
    \caption{Proposed experimental setup using photon pairs from SPDC to produce heralded single photons. See more details in the main text.}
    \label{setup}
\end{figure}

The experimental scheme just described can be implemented with commercially available devices. Fig. \ref{setup} shows a proposed experimental setup that could run the protocol. It uses a 405~nm continuous wave (c.w.) diode laser to pump a nonlinear crystal and  produce pairs of photons trough SPDC. Suppose that one uses a type II phase matching interaction with a beta barium borate (BBO) crystal so that signal and idler photons are emmitted with orthogonal polarizations, both at 810~nm wavelength and they are separated wth a polarizing beam splitter (PBS). One of the photons is detected by a Single Photon Counting Module (SPCM) at Alice's station. This detection event heralds its twin photon that is transmitted to the Bob's station using an optical system composed by an image or Fourier lens and a telescope. The photonic optical mode is also prepared to encode one character of the message through its position in the transverse spatial distribution as explained before. This encoding can be made mechanically by simply using a rotating disk with one aperture. However, a faster modulation can be made using a Digital Micro Mirror Device (DMD).

The modulated photon is received at Bob's station using a telescope link and one image or Fourier lens. The photonic mode is detected with a photon counting matrix device, which is gated by a pulse produced by the detection of the Alice's heralding photon. 

This procedure has the following advantages: i) the heralding scheme mitigates photon loss effects, because only events resulting from a coincidence between the heralding photon and the matrix detected photon are accounted; ii) it also allows the timing control, because the detection of the heralding photon generates a time tag. Time tagging control is essential for the protocol, because the Reed-Solomon protocol uses packages of ten sequential photonic characters; iii) matrix photon counting is a technology that is being developed very fast due to its several applications. Therefore, one has a few available technologies that can be used like matrices of APDs, i-CCM and i-CMOS cameras with built-in time taggers. 

We conclude that the protocol can be implemented with current technology available commercially. 

\section{Discussion and security analysis}

\begin{figure}[!htb]
    \centering
    \includegraphics[width=1\linewidth]{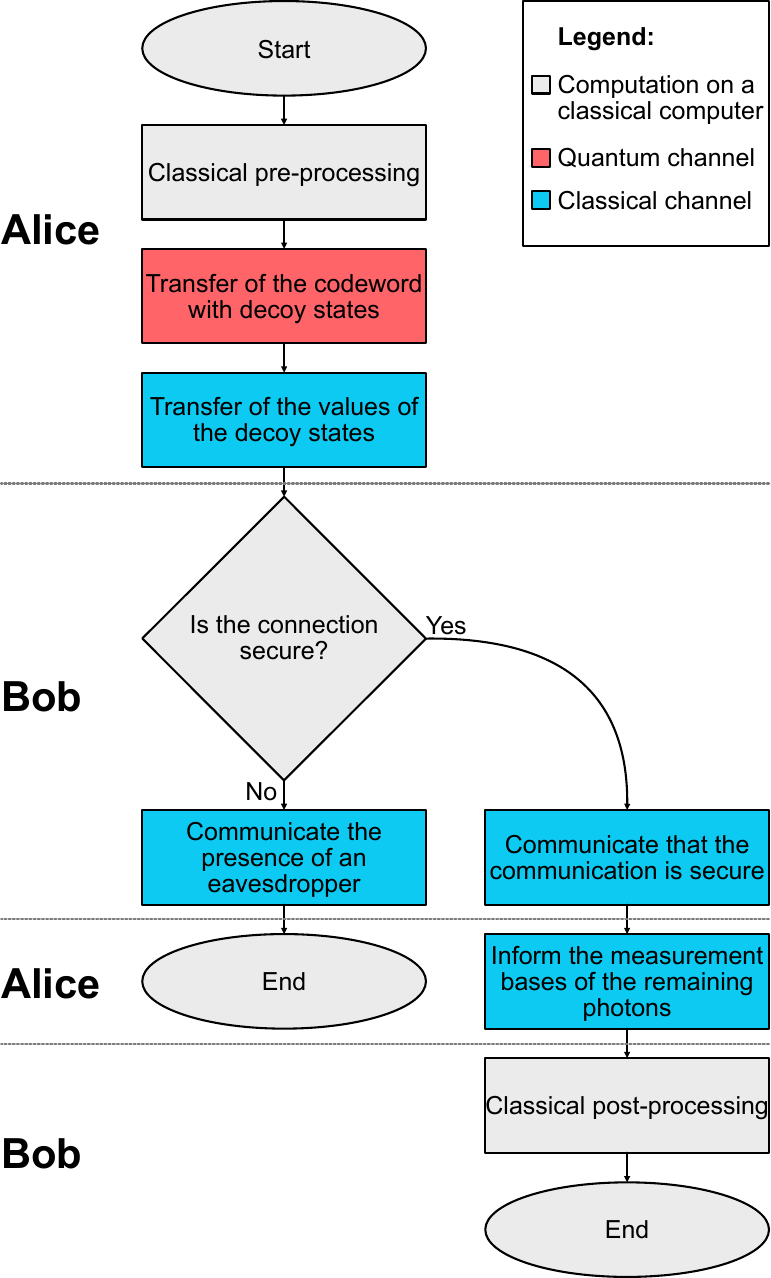}
    \caption{Flowchart of the protocol}
    \label{fig:flowchart}
\end{figure}

The flowchart of the protocol is illustrated in Figure~\ref{fig:flowchart}. The advantages of this scheme are: its simplicity, relying mostly on classical resources, operating on already existing technology, leaving quantum communication restricted to the most sensible data, and not requiring quantum memory.

The use of decoy states is an important piece, leading to a decision of whether the communication should continue or not, and making the protocol resilient against intercept-resend attacks. In a more general form, when Eve intercepts a photon and resends it to Bob, she adds an error rate equivalent to the probability of measuring the photon on the wrong basis and preparing the wrong character. For this scheme, the number of MUBs is $b = 2$ so the error introduced by Eve can be calculated as in Equation~\ref{eq:eve_ir}.

\begin{equation}\label{eq:eve_ir}
    E_{IR} = \frac{b-1}{b} \frac{d-1}{d} \implies E_{IR} = \frac{1}{2} \frac{d-1}{d}
\end{equation}

For $b = 2$, the the error introduced by Eve is limited to 50\%. The use of $b > 2$ increases the introduced error, but has the drawbacks of being physically more difficult to find systems with more MUBs operating in high dimensions and would also require the transmission of more parity photons, since Bob would pick the wrong basis more often.

\section{Conclusion}

In this work, we have proposed a novel and highly viable Deterministic Secure Quantum Communication (DSQC) protocol that mitigates the critical challenges of channel loss and low bit-rate by integrating classical Reed-Solomon erasure coding with high-dimensional quantum states. By encoding information in the transverse spatial degrees of freedom (TSDF) of single photons, the protocol inherently expands the channel capacity beyond standard binary quantum key distribution schemes.  A primary advantage of the proposed architecture is its reliance on commercially available technology and classical pre- and post-processing, entirely bypassing the current technological bottleneck of quantum memory. We demonstrated that the transmission of a deterministic message can be mathematically protected against network noise through block-wise polynomial evaluation, while simultaneously remaining resilient to intercept-resend eavesdropping attacks via the integration of decoy states and mutually unbiased bases (MUBs). Furthermore, our proposed experimental implementation—utilizing spatial light modulators, heralded single photons from spontaneous parametric down-conversion, and modern APD matrix detectors—confirms that this theoretical framework is immediately adaptable to current laboratory capabilities.  Ultimately, leveraging structured light to multiplex high-dimensional quantum states with robust classical erasure codes opens a highly pragmatic pathway for deploying secure, high-capacity quantum communication networks in noisy, real-world environments.

\begin{acknowledgements}
This work has been supported by the Brazilian agencies CNPq
(DOI 501100003593), CAPES (DOI 501100002322), FAPESC
(DOI 501100005667, DOI 2025TR001683),  Call 62/2024, and INCT IQNano (406636/2022 2), INCT DQ (408783/2024 9) and  FAPEG (202510267001843).
\end{acknowledgements}

\end{document}